# Uniform non-stoichiometric titanium nitride thin films for improved kinetic inductance detector array

G. Coiffard[1,2] • K-F. Schuster[1] • E.F.C. Driessen[1] • S. Pignard[2] • M. Calvo[3] • A. Catalano[4,3] • J. Goupy[3] • A. Monfardini[3,4]

[1] IRAM, Institut de Radioastronomie Millimétrique, 300 rue de la Piscine, 38406 S$^t$ Martin d'Hères, France
[2] Univ. Grenoble Alpes, LMGP, F-38000 Grenoble, France
[3] CNRS & Univ. Grenoble Alpes, Institut Néel, F-38042 Grenoble, France
[4] Laboratoire de Physique Subatomique et de Cosmologie, CNRS/IN2P3, UJF, INP, Grenoble, France

**Abstract** We describe the fabrication of homogeneous sub-stoichiometric titanium nitride films for microwave kinetic inductance detector (KID) arrays. Using a 6" sputtering target and a homogeneous nitrogen inlet, the variation of the critical temperature over a 2" wafer was reduced to <25 %. Measurements of a 132-pixel KID array from these films reveal a sensitivity of 16 kHz/pW in the 100 GHz band, comparable to the best aluminium KIDs. We measured a noise equivalent power of NEP = $3.6\times10^{-15}$ W/Hz$^{1/2}$. Finally, we describe possible routes to further improve the performance of these TiN KID arrays.

## 1 Introduction

Sub-stoichiometric titanium nitride (TiN$_x$) has been demonstrated to work as an alternative material for microwave kinetic-inductance detectors (KIDs) [1-4]. TiN offers two main advantages over more traditional aluminum. Firstly, the critical temperature of the TiN is tunable with the nitrogen content, $0.5 < T_C < 4.5$ K, which allows for lower frequencies (down to 36 GHz) to be detected, whereas aluminum is limited to frequencies above 105 GHz.
Secondly, the large normal-state resistivity of TiN ($\rho > 80$ μΩcm) allows for an easy design of lumped-element KIDs [5] and a compact pixel size, due to the equally increased kinetic inductance.

G. Coiffard • K-F. Schuster • E.F.C. Driessen • S. Pignard • M. Calvo • A. Catalano • J. Goupy • A. Monfardini

Due to the very strong variation of $T_C$ with nitrogen content, however, large scale uniformity of the film composition becomes an issue when large arrays of detectors are desired, such as for the NIKA 2 instrument [6]

In this paper, we describe an upgrade of our deposition chamber that improves the uniformity of our TiN films from 50 % to 75 %. We show that this better uniformity allows us to create large TiN KID arrays with sensitivity close to the best reported aluminum KID arrays, but with still an increased noise level. Finally, we describe avenues to improve the performance of our TiN arrays.

## 2 TiN thin films deposition and characterization

2.1 Configuration of the deposition chamber

The TiN films presented in this paper were deposited using magnetron sputtering in a Plassys MP500s [7,8]. We compare three different configurations of the deposition chamber, as shown in Fig. 1. In the first configuration, a 4'' Ti target was used combined with a single-point gas injection for the Ar and $N_2$ gases. In the second configuration, a 6'' Ti target was used, with a single-point gas injection. In the third one, we use a 6'' target and a ring injection system that allows for a more uniform application of the $N_2$ gas. In all three configurations, the distance between the target and the non-rotating substrate was 10 cm.

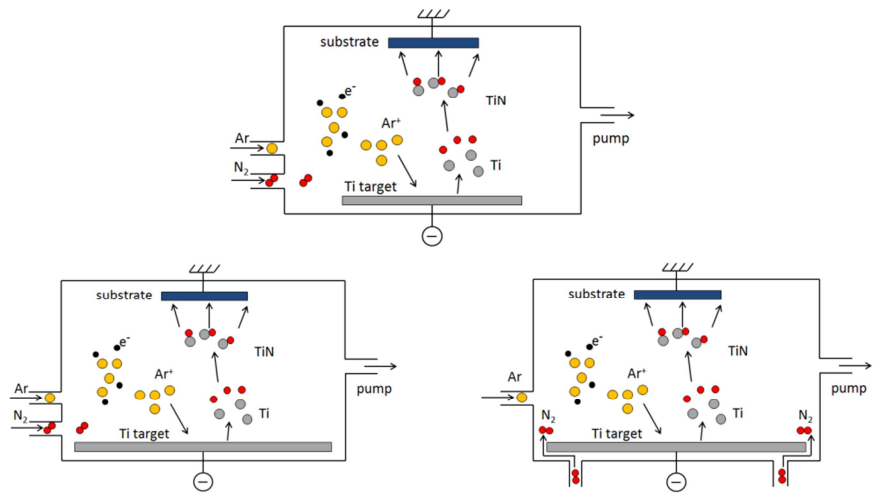

**Fig. 1** *Top* Illustration of the deposition chamber with the 4'' Ti target and single-point injection *Bottom left* 6'' target and single-point gas injection *Bottom right* 6'' target and ring injection for $N_2$ (color figure online)

**Uniform non-stoichiometric titanium nitride thin films for improved kinetic inductance detector array**

2.2 Deposition process

Prior to any deposition, pure Ti is sputtered in the empty deposition chamber (base pressure $2\text{-}3\times10^{-8}$ mbar) in order to trap contaminant species (mostly $H_2O$ and $O_2$). The quantities of the different species present in the reactive chamber are monitored with a mass spectrometer. After this, the substrate is moved from the loadlock to the chamber and cleaned with an Ar RF plasma. Then, with the shutter closed, the Ti target is pre-sputtered for a few minutes to ensure a sputtering of pure Ti. The Ar flow is set to 30 sccm, the process pressure is fixed to 0.2 Pa and 1000 W are delivered to the target. Still with the shutter closed, the $N_2$ valve is opened and the flow is chosen in order to achieve different stoichiometries. After approximately 3 min, the shutter is opened and the desired thickness of TiN is sputtered on a 2'' silicon substrate. Many films were deposited under various $N_2$ flow rates. The samples spread from pure Ti film with $N_2$ flux set to zero to stoichiometric TiN. Each TiN sample was cut into pieces for ellipsometry and superconducting transition temperature measurements. Some films were mapped and others were used to fabricate KID arrays.

2.3 Characterization

The critical temperature of various samples was measured using a four-point resistance measurement in a $^4$He cryostat equipped with a $^3$He sorption cooler. As observed earlier by Vissers et al. [9], the measured critical temperature is correlated to the ellipsometry parameter Δ, which was measured with a Sentech SE400 ellipsometer at a wavelength of 632.8 nm and an angle of incidence of 70 degrees (Fig. 2 *top left*). Using this correlation allowed us to map the critical temperature of a complete wafer at room temperature. This method allows us to map local differences in Tc with an estimated accuracy of 4 mK.

Fig. 2 *top right* presents values of $T_C$ versus the radial position on 2'' TiN substrates prepared with the three different configurations of the deposition chamber. Firstly, we greatly improved the uniformity defined as the ratio of lower $T_C$ and higher $T_C$ by changing the size of the Ti target; then the ring injection allows to further improve the uniformity in $T_C$. As the $T_C$ distribution is not concentric, it appears more clearly in the mapping in Fig. 2. The uniformity in $T_C$ is improved from 50% to 75% with the use of a ring injection. In Fig. 2 *bottom*, the $N_2$ inlets are shown. In the single point injection configuration, the critical temperature in the film is higher near the gas injection point and the further from the inlet, the lower becomes $T_C$. It is

G. Coiffard • K-F. Schuster • E.F.C. Driessen • S. Pignard • M. Calvo • A. Catalano • J. Goupy • A. Monfardini

interpreted as being due to the depletion of the nitrogen in the plasma. With the ring injection, we see a similar effect. TiN films are deposited under a low $N_2$ flow rate and it seems that the $N_2$ is not equally distributed around the target, consistent with the position of the $N_2$ injection point into the ring, on the top right. Vissers *et al.* [9] have reached uniformity around 75 % on 3 inch rotating substrate. Here, we obtain comparable results which we were able to link to the nitrogen injection.

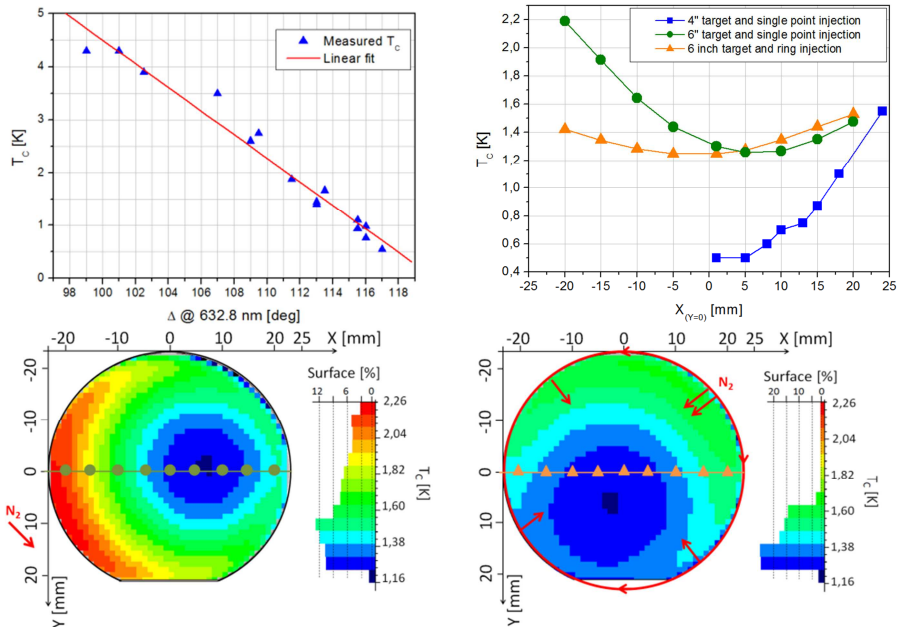

**Fig. 2** *Top left* The critical temperature $T_C$ scales with the optical parameter $\Delta$ determined with the ellipsometer *Top right* Value of $T_C$ vs. the radial position on the wafer for three different configuration of the deposition chamber *Bottom left* Mapping of the $T_C$ of TiN substrate with 6'' Ti target and the single point injection *Bottom right* Ring injection. The $N_2$ inlets for both configuration are shown (color figure online)

**3 KID arrays**

3.1 Design and experimental setup

To assess the usability of our improved TiN films for millimeter wave detection, a 50-nm TiN film with a $T_C$ of 0.8 K and a uniformity of 75 % was selected. The array counts 132 LEKID resonators (size 2x2 mm$^2$); all of them are connected to a single CPW transmission line and they are designed

**Uniform non-stoichiometric titanium nitride thin films for improved kinetic inductance detector array**

to resonate between 1 and 2 GHz. The 36x36 mm$^2$ array is structured by optical lithography and is etched by Reactive Ion Etching in a mixture of $SF_6/C_4F_8$, using interferometric end point detection.

We added wire bonds across the CPW feedline to prevent propagation of parasitic modes [10]. The array is mounted and connected in a cryostat designed for electrical and optical measurements at a base temperature of 100 mK. As shown in Fig. 4, the array is back-side illuminated through the Si substrate and a backshort placed at a calculated distance allows to maximize absorption in the array at 140 GHz. We also added a low-pass filter at 180 GHz in the optical path to analyze the response between the superconducting gap frequency and the filter cutoff.

3.2 Electrical and optical measurements

The transmission of the microwave feedline of the array was measured using a vector network analyzer. We identified 124 resonances between 1.3 and 1.7 GHz (see Fig. 3) which corresponds to a specific kinetic inductance of $L_k$ = 20 pH/sq. Comparing to simulations, we find a max shift of $L_{kin}$ of 10%. The responsivity of the array was determined using two different black-body sources placed in front of the cryostat windows, equivalent to a temperature of 300 K and 200 K. The average measured frequency shift was measured to be 48.4 ± 12 kHz corresponding to a response of 16.1 ± 8.8 kHz/pW at the FWHM (120-150 GHz) of the absorption peak seen in Fig. 4. In this band, the variation of the incident power seen by the detector is determined with an optical simulation of the cryostat [10]. The frequency noise was measured to be 58.6 ± 12 Hz/Hz$^{1/2}$. This translates into an average noise equivalent power (NEP) of 3.6 ± 2.7×10$^{-15}$ W/Hz$^{1/2}$. In table 1, these numbers are compared with the one of aluminum detectors for NIKA; the data for our TiN detector in the full available bandwidth from 65 to 180 GHz are equally presented. Our TiN array shows poor performances especially due to the very high detector noise. The noise spectrum at our readout frequency of 24 Hz is limited by a 1/f behavior usually attributed to two-level system noise [1]. Therefore, an improved NEP could be obtained by moving to higher readout frequencies. The optical response is comparable to the best Al arrays however.

G. Coiffard • K-F. Schuster • E.F.C. Driessen • S. Pignard • M. Calvo • A. Catalano • J. Goupy • A. Monfardini

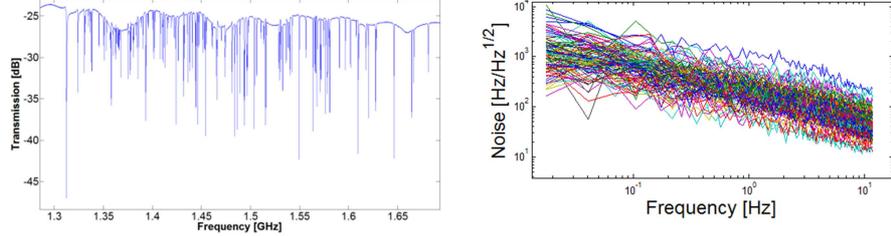

**Fig. 3** *Left* VNA scan of a 132-resonator array. Each dip corresponds to a KID resonator *Inset*. Response of 2 KIDs under two different optical loadings *Right* Noise spectrum for all KIDs (color figure online)

|  | Al actual band | TiN full band | TiN actual band |
|---|---|---|---|
| Response [kHz/pW] | 16 | 4 ± 2 | 16.1 ± 8.8 |
| Noise [Hz/Hz$^{1/2}$] | 1 | 58 ± 12 | 58.6 ± 12 |
| NEP [W/Hz$^{1/2}$] | 6×10$^{-17}$ | 1.4 ± 1 ×10$^{-14}$ | 3.6 ± 2.7×10$^{-15}$ |

**Table 1.** Comparison of performances of Al [10] and TiN KID array. The response in the actual band is at the FWHM of the absorption spectra. The full band lies from the superconducting gap frequency to the filter cutoff.

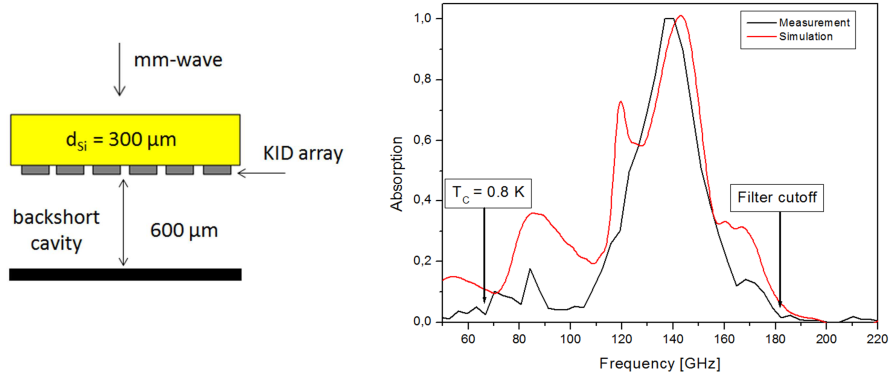

**Fig. 4** *Left* Illustration of the structure made of the Si substrate, the TiN KID array and the backshort. The array is back-side illuminated *Right* Absorption spectrum of the TiN array. This spectrum corresponds to the average of 14 pixels. The cutoff appears at 65 GHz, consistent with the estimated $T_C = 0.8$ K according to the relationship $h\nu = 4k_B T_C$ [11] (color figure online)

We have also measured the frequency-dependent optical absorption of our array using a Martin-Puplett interferometer placed in front of the cryostat

**Uniform non-stoichiometric titanium nitride thin films for improved kinetic inductance detector array**

window [12]. Fig. 4 shows the average normalized absorption of 14 pixels. It can be seen that the maximum absorption occurs in a much narrower band than what is defined by the superconductivity gap frequency (60 GHz) and the low-pass filter (180 GHz). We also performed finite element simulation of the geometry of the entire KID including the feedline, the ground plane, the inductor and the interdigital capacitor. The structure is made of the substrate, the KID and the backshort and a mm-wave port illuminates it through the substrate. The absorption of the whole structure is then derived from the signal reflected into the port. The simulated absorption (red curve in Fig. 4) reveals that this spectral behavior is largely due to the not optimized geometry of pixel and backshort.

**4 Conclusion**

The value of the superconducting transition temperature is estimated with an ellipsometry measurement that allows TiN thin films to be mapped to visualize the uniformity. The uniformity of sputtered TiN thin films is improved with a larger Ti target; and thanks to a $N_2$ ring injection we were able to further improve the uniformity by a factor of two. A better uniformity can still be reached by further improving the nitrogen repartition in the process chamber. The first TiN array measured shows a responsivity still low and a noise relatively high compared to Al-KID. The responsivity can still be improved with optimized pixel geometry, backshort distance, a smaller volume of the superconductor and antireflection layer [6]. This way, KID arrays that allow detection of radiation at frequencies < 105 GHz come into reach.